\documentclass[aps,pre,twocolumn,groupedaddress]{revtex4-1}

\usepackage{amsmath,amssymb}

\usepackage{color}
\usepackage{graphicx}
\usepackage{soul}

\usepackage{natbib}

\begin{document}

\title{On the analogy of quantum wave-particle duality with bouncing droplets}
\author{Chris D. Richardson}
\author{Peter Schlagheck}
\author{John Martin}
\author{Nicolas Vandewalle}
\author{Thierry Bastin}
\affiliation{D\'epartement de Physique, University of Liege, 4000 Liege, Belgium}
\date{\today}

\begin{abstract}
We explore the hydrodynamic analogues of quantum wave-particle duality in the context of a bouncing droplet system which we model in such a way as to promote comparisons to the de Broglie-Bohm interpretation of quantum mechanics.  Through numerical means we obtain single-slit diffraction and double-slit interference patterns that strongly resemble those reported in experiment and that reflect a striking resemblance to quantum diffraction and interference on a phenomenological level.  We, however, identify evident differences from quantum mechanics which arise from the governing equations at the fundamental level.

\end{abstract}

\maketitle

\section{Introduction}

Hydrodynamic analogues of quantum wave-particle duality have attracted considerable attention during the past decade~\cite{bib:CouderSingle,Bush12102010,CouderWalkingNature,FLM:431290,1742-6596-361-1-012001,refId0,PhysRevLett.102.240401,0295-5075-102-1-16005,FLM:9203453,FLM:9131969,Fort12102010,FLM:8266690}. This started with the relatively recent discovery of the phenomenon of a ``walker''~\cite{CouderWalkingNature,FLM:431290}, formed by a droplet that is periodically bouncing above the surface of a vertically vibrated liquid. In a suitable regime of the parameters that characterize the vertical vibration, this droplet starts to move horizontally as a consequence of its interaction with the surface waves that it creates during the bouncing. While this horizontal motion is uniform in a homogeneous liquid, it becomes significantly perturbed in the vicinity of boundaries or sub-surface obstacles, due to total or partial reflection of the emitted surface waves, giving rise to irregular trajectories of the droplet. In a pioneering experiment conducted by Couder and Fort~\cite{bib:CouderSingle}, such droplets were individually launched at obstacles that contained one or two slits across which the droplets could pass. Monitoring the positions of the droplets that had managed to pass this single or double-slit configuration yielded similar single-slit diffraction and double-slit interference patterns as the ones well known from quantum mechanics~\cite{bib:CouderSingle,1742-6596-361-1-012001}. Subsequent experiments yielded other hydrodynamic analogs of quantum phenomena, such as tunneling~\cite{PhysRevLett.102.240401}, orbit quantization~\cite{0295-5075-102-1-16005,FLM:9203453,FLM:9131969}, and Landau levels~\cite{Fort12102010}.

While there exist many classical analogs of quantum wave interference phenomena, in particular in the electrodynamic context (such as Anderson localization~\cite{wiersma1997localization,wiersma1997localization,PhysRevLett.96.063904} and weak localization~\cite{407,PhysRevLett.55.2696} in disordered systems), walking droplets are recognized as the first and, to our knowledge, still unique classical analog of quantum \emph{wave-particle duality}. Indeed, in the context of the double-slit experiment of Ref.~\cite{bib:CouderSingle} individual droplets are observed to walk through only one of the two slits within the obstacle, and yet the interference fringes observable in their final probability distribution clearly reflect that the motion of the droplets was influenced by the presence of the other slit.  This is very similar to quantum interference experiments with molecules~\cite{bib:arndt} although in this case the entire trajectories of the particles can not be observed without affecting the final probability distribution.  It is thought that the reason for the interference resulting from the trajectories of many single droplets is that, unlike the droplet itself, the droplet's associated surface wave passes through both slits, and it is this subsequent wave which guides the droplet into a trajectory which results in interference.  These phenomena of a particle (droplet) as a source of waves and that particle being \emph{piloted} by a wave (not necessarily originating from the particle) is strongly reminiscent of de Broglie's two solutions interpretation of quantum mechanics~\cite{de1926ondes} which was later reformulated by Bohm~\cite{PhysRev.85.166,PhysRev.85.180} into a usable theory which is expected to reproduce all of the features of quantum mechanics.

From a quantitative point of view, however, this novel classical analogy of quantum wave-particle duality in the bouncing droplet system is not at all evident, and there are many arguments that one can raise against it.  While Bohmian quantum mechanics exhibits nonlocal features~\cite{PhysRev.85.166,PhysRev.85.180,durr2010bohmian,sep-qm-bohm}, the evolution of the droplet and the surface waves is rooted in hydrodynamics which is manifestly a local theory, unless incompressibility is assumed.  In the de Broglie-Bohm interpretation, the specific trajectory of the quantum particle does not back-act onto the evolution of the wavefunction, whereas the droplet creates new surface waves at the position where it bounces. Those surface waves do evolve, to a very good approximation, according to a linear theory, but a direct mapping to the Schr\"odinger equation is not obvious.  The force on a quantum particle in the de Broglie-Bohm interpretation flows directly from the Schr\"odinger equation and so lack of an equivalent mapping in the droplet system will also lead to a lack of an equivalent mapping to the force on a droplet.  In view of these concerns, it may even appear surprising that so many close analogies with quantum mechanics were reported in the above-cited droplet experiments~\cite{bib:CouderSingle,Bush12102010,CouderWalkingNature,FLM:431290,1742-6596-361-1-012001,refId0,PhysRevLett.102.240401,0295-5075-102-1-16005,FLM:9203453,FLM:9131969,Fort12102010,FLM:8266690}.

The main purpose of this paper is to elaborate, on the basis of numerical simulations, why and to which extent this striking analogy exists between the classical droplet system and the behavior of a quantum particle as far as the \emph{observable phenomenology} is concerned. We shall particularly focus on single-slit diffraction and double-slit interference as in the experiments of Ref.~\cite{1742-6596-361-1-012001}.  We shall make use of theories for the motion of the droplet and the evolution of the surface wave that have proved their validity in an experimental context~\cite{FLM:431290,bib:CouderSingle,CouderWalkingNature,FLM:9203453,FLM:8266690}. We shall, moreover, consider several idealizations (such as a very large "memory"~\cite{FLM:8266690} of the surface wave as far as previous bounces of the droplet are concerned) and discuss to which extent one can relax those constraints while still observing quantum-like phenomena with good visibility.

The paper is organized as follows: in Sec.~\ref{sec:model} we describe an idealized hydrodynamic model of the bouncing droplet system and compare some features of the model to quantum mechanics.  We first, in Sec.~\ref{sec:force}, characterize the force on a droplet and explore the analogy of this hydrodynamic force and wave to a Bohmian pilot wave and quantum force.  We then, in Sec.~\ref{sec:walker}, describe the surface wave excited by a bouncing walking droplet.  In Sec.~\ref{sec:slits} we propose a possible mechanism underlying the diffraction and interference of the single and double-slit experiments of Ref.~\cite{bib:CouderSingle} by using the surface wave from Sec.~\ref{sec:walker} as an input to the slit configuration and the force equation from Sec.~\ref{sec:force}.  We conclude in Sec.~\ref{sec:conclusion}.

\section{The hydrodynamic model} \label{sec:model}

Here we develop a model for the bouncing droplet system that will serve as a base for comparisons to quantum theory.  The droplet and the wave it excites are both macroscopically distinct entities and must be described separately.  The droplet experiences a newtonian force and the wave, which is updated by the bouncing of the droplet,  obeys a wave equation.   We can draw parallels from this to quantum mechanics using the de Broglie-Bohm interpretation of quantum mechanics~\cite{PhysRev.85.166} in which a quantum particle is understood as a point-like object absent of all properties besides position.  Its motion is guided by a quantum force derived from the wave function, $\psi$, associated with the particular experiment and which evolves according to the standard Schr\"odinger equation.  Therefore a description of the forces on the droplet and the surface wave in the bath will allow direct comparisons to the quantum fore on a particle and the wave function in the de Broglie-Bohm interpretation of quantum mechanics.

\subsection{The force on a droplet and potential analogues}  \label{sec:force}

It is well understood~\cite{FLM:431290,bib:CouderSingle,CouderWalkingNature} that the droplet's velocity between bounces will change in proportion to the angle of the surface wave at the point where the droplet impacts it.  We can therefore follow Couder and Fort~\cite{1742-6596-361-1-012001} and relate the change of velocity, due to an impact with the bath, to the gradient of the surface wave immediately before the impact and at the point of impact as
\begin{align}
\Delta \dot{\mathbf{r}} &\propto -  \nabla \zeta(\mathbf{r},t) , \label{eqn:CoudersPot}
\end{align}
where $\zeta(\mathbf{r},t)$ is the height of the surface wave with positive values opposite the direction of gravity.  Taking into account that the acceleration $\ddot{\mathbf{r}}$ is proportional to the change in velocity $\Delta \dot{\mathbf{r}}$, and adding a viscous damping term, $f^v$, the average Newtonian force on the droplet over a bouncing period takes the simple form
\begin{align}
m \ddot{\mathbf{r}} &=  - \Gamma^b \nabla \zeta(\mathbf{r},t) - f^v \dot{\mathbf{r}} , \label{eqn:DiscreteForce}
\end{align}
where $m$ is the mass of the droplet, and $\Gamma^b$ is a force proportionality constant.

Typically in experiment the mass of the droplet is on the order of $m \sim 1\,\mathrm{mg}$~\cite{bib:CouderSingle}.  We estimate the maximum value for the magnitude of the total force to be on the order of $10^{-6}\,\mathrm{N}$~\cite{FLM:431290}.  Using a maximum surface wave amplitude of $0.01\,\mathrm{mm}$ and a typical Faraday wavelength of $\lambda_F \approx 6\,\mathrm{mm}$~\cite{bib:CouderSingle} we estimate the maximum value for the gradient of the surface wave, $\nabla \zeta(\mathbf{r})$, to be approximately $10^{-2}$, leading to a value for the force pre-factor $\Gamma^b \sim 10^{-4}\,\mathrm{N}$.  In addition, the viscous damping term is on the order of $f^v \sim 10^{-6}\,\mathrm{N}\,\mathrm{m}^{-1}\,\mathrm{s}$~\cite{FLM:431290}, and must be accounted for in order to avoid unrealistically large velocities for the droplet. 

\begin{figure}[ht]
\includegraphics[width=1\linewidth]{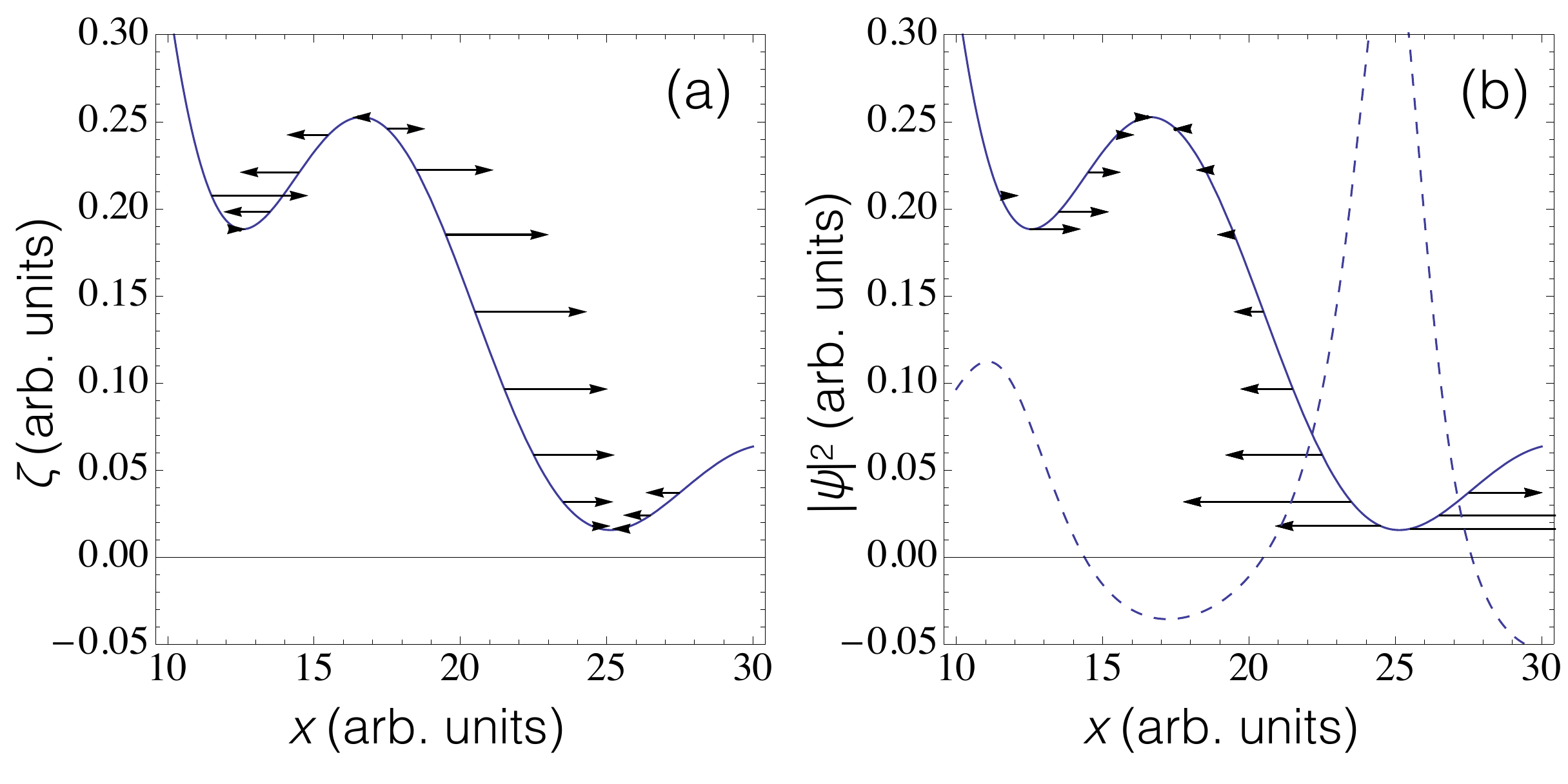}
\caption{(a) The arrows depict the relative magnitudes and direction of the force on a droplet calculated from the conservative part of the hydrodynamic force, Eq.~(\ref{eqn:DiscreteForce}), along an arbitrary surface wave (solid line).  (b) The arrows depict the relative magnitudes and direction of the force on a quantum particle calculated from Bohm's quantum potential, Eq.~(\ref{eqn:qmp}), along an probability density (solid line) with the same form as in (a).  The dashed line is Bohm's quantum potential in arbitrary units.  Note that at the points where the probability is zero the force on the particle is infinite.  Note that the droplet is encouraged to move towards the trough of the wave and the quantum force is singular at zero probability.  The forces derived from the wave function and the surface wave are fundamentally different but can still lead to a probabilistic analogy.}
\label{fig:ForceComparison}
\end{figure}

We can now explore a possible analogy between the excited surface wave $\zeta$ and the quantum wave function $\psi$.  To do so we formulate the action of the wave function on the particle, by way of the Madelung transformation~\cite{bib:madelung}, into a Newtonian potential, Bohm's quantum potential~\cite{PhysRev.85.166,PhysRev.85.180},
 \begin{align}
U(\mathbf{r},t) &= - \frac{\hbar^2}{2 m} \frac{\nabla^2 \left| \psi(\mathbf{r},t) \right|}{\left| \psi(\mathbf{r},t) \right|} , \label{eqn:qmp} 
\end{align}
which inserted into Newton's equation,
\begin{align}
m \ddot{\mathbf{r}} &= - \nabla U(\mathbf{r},t) , \label{eqn:quantforce}
\end{align}
gives rise to all the uniquely quantum behavior seen in quantum systems.  To demonstrate the differences between this quantum force derived from the wave function and the hydrodynamic force from Eq.~(\ref{eqn:DiscreteForce}) we explore a one dimensional slice of a quantum probability density shown in Fig.~\ref{fig:ForceComparison}(b) and an identical slice of a surface wave in Fig.~\ref{fig:ForceComparison}(a).  We visualize the magnitude and direction of the force on the droplet using the conservative part of Eq.~(\ref{eqn:DiscreteForce}) and on the quantum particle using Bohm's quantum force given in Eq.~(\ref{eqn:quantforce}).  We can immediately see from Fig.~\ref{fig:ForceComparison} that there is no mapping between the forces resulting from the quantum wave function $\psi$ and the height of the surface wave $\zeta$.  Indeed, looking at the behavior of the quantum force, Fig.~\ref{fig:ForceComparison}(b), whenever the probability density is close to zero, the quantum force becomes singular and will very quickly remove any particle in the area.  Conversely, the hydrodynamic force directs the droplet into the trough of the wave as can be seen in Fig.~\ref{fig:ForceComparison}(a).  The effect of the the forces from the two guiding equations is analogous if we consider those areas of the effective surface wave with large slope to be equivalent to areas of low probability in the quantum wavefunction.  Both forcing equations apply a strong force that will remove a particle from an area of low probability (high slope).  Further, because the force and likely therefore the velocity of the droplet in an area of low probability (high slope) is large, the droplet will spend less time in that area and be less likely to be found there.  As it stands, we should instead define the surface wave of the droplet as a \emph{potential} $U_{\textrm{d}} \propto \zeta$, which then permits direct comparison to Bohm's quantum potential $U$ (dashed line in Fig.~\ref{fig:ForceComparison}(b)).  It can then be noticed that the two potentials that give rise to the quantum and hydrodynamic forces forces are relatively dissimilar.

\subsection{The excited wave of a walker}  \label{sec:walker}

Consider a walking droplet bouncing on a vibrating surface with a constant droplet speed on the order of $10 \,\mathrm{mm/s}$~\cite{bib:CouderSingle}, which is on the order of one tenth the phase velocity of an excited wave~\cite{bib:CouderSingle}.  Due to the low velocity of the droplet compared to the phase velocity of the Faraday wave it excites, we can approximate the excited Faraday wave due to a bounce of a droplet as a spatially non-local wave.  Because of the energy imparted to the bath by the vibrator, this Faraday wave excited by the bounce of the droplet is made to persist with a lifetime $\tau$.  Therefore, the non-local wave that builds up after $N$ bounces of the droplet is a sum of standing waves with temporally decaying amplitudes given by~\cite{FLM:8266690,0295-5075-102-1-16005}
\begin{align}
\zeta_{N}(\mathbf{r},t) &= A \sum\limits_{n=1}^N \cos(\omega_F t)  J_0\left(k_F |\mathbf{r} - \mathbf{r}_n | \right) e^{-(t_N - t_n) / \tau} .\label{eqn:damped_intitial}
\end{align}
Here $\zeta_{N}$ is the amplitude of the excited surface wave after $N$ bounces, $A$ is the amplitude after one bounce, $t_n$ is the time of the $n$th bounce, $t_N$ is equal to the total evolution time of the surface wave, $\omega_F$ is the angular frequency of the surface wave,  $k_F$ is its wave number and is on the order of $1 \,\mathrm{rad/mm}$, $J_0$ is the Bessel function of the first kind of order $0$, and $\mathbf{r}_n = \mathbf{r}_0 - n \Delta \mathbf{r} = \mathbf{r}_0 - \mathbf{v}_d t_n$ is the position of the droplet at the $n$th bounce with $\mathbf{v}_d$ being the constant velocity of a walking droplet and $\mathbf{r}_0$ the position of the droplet at the first bounce.

The lifetime $\tau$ of an excited surface wave increases with increasing amplitude of the vibrator and will theoretically increase indefinitely the closer the driving is tuned to the Faraday threshold, the point at which the surface becomes unstable.  Because we are dealing with a known characteristic bouncing period, the Faraday period $T_F$, it is more convenient to express the lifetime $\tau$ in terms of it.  Eddi \emph{et al.}~\cite{FLM:8266690} introduced the dimensionless memory parameter $M \equiv \tau / T_F$ which is equivalent to the number of bounces of the droplet the bath will store.  This memory parameter permits us to transform the temporal damping function in Eq.~(\ref{eqn:damped_intitial}) into the form $\exp(-(t_N - t_n) / \tau) = \exp(-(N - n) / M)$ and to expand the parameters $t_n = n T_F$, $t_N = N T_F$, and $\omega_F = 2 \pi / T_F$.

\begin{figure}[ht]
\includegraphics[width=1.0\linewidth]{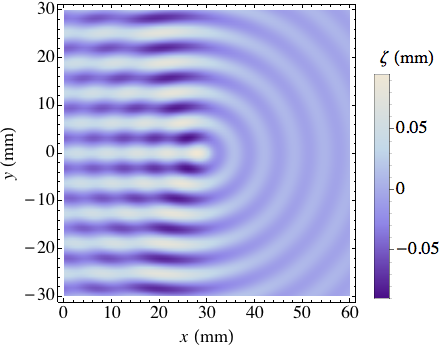}
\caption{(color online) A visualization of the surface wave, Eq.~(\ref{eqn:exactsum}), created by a walking droplet traveling to the right with a speed $v_d = 10\,\mathrm{mm/s}$ with the memory parameter $M = 100$, Faraday period $T_F = 0.04\,\mathrm{s}$, and wavenumber $k_F=1\,\mathrm{rad/s}$.  The surface wave is a superposition of excited Faraday waves due to many bounces of a walking droplet moving with a constant velocity.}
\label{fig:WaveFront3D}
\end{figure}

The walking droplet bounces synchronously with the vibrator~\cite{CouderWalkingNature} and therefore only interacts with the surface wave \emph{stroboscopically}.  From one bounce to the next one time is incremented by $\Delta t = T_F =2 \pi / \omega_F$ and so we can restrict the consideration to those times at which the droplet interacts with the bath, $t_n = n \Delta t $.  We can, therefore, drop the value of the explicitly time dependent part of Eq.~(\ref{eqn:damped_intitial}) as $\cos(\omega_F t_n) = 1$, which results in
\begin{align}
\zeta_{N}(\mathbf{r}) =  A \sum\limits_{n=1}^N J_0\left(k_F |\mathbf{r} - \mathbf{r}_n | \right) e^{-(N - n) / M} .  \label{eqn:exactsum}
\end{align}
This is a similar equation as the the one given in~\cite{FLM:8266690,Fort12102010} where we differ by not using the asymptotic expansion of the Bessel function and neglecting the spatial damping term due to the viscosity of the fluid.  The horseshoe shaped surface wave depicted in Fig.~\ref{fig:WaveFront3D} is typical of a walker and is commonly seen in experiment~\cite{0295-5075-102-1-16005,FLM:8266690}.  As can be seen the shape of the wave behind the droplet is quite complex, but as we are studying diffraction and interference, we note that it is only the front of the wave which will impinge on a slit as the droplet approaches the slit.  We can, therefore, disregard the complexity and focus only on the forward part of the surface wave.

We consider the portion of the excited surface wave that impinges on a slit when the droplet is still a large distance away from those slits.  In this limit, 
\begin{align}
|\mathbf{r} - &\mathbf{r}_n| =  |\mathbf{r} - \mathbf{r}_N + (N-n) \Delta \mathbf{r}| \\
&\approx |\mathbf{r} - \mathbf{r}_N| + (N-n) \Delta \mathbf{r} \cdot \frac{\mathbf{r} - \mathbf{r}_N}{|\mathbf{r} - \mathbf{r}_N|} + \mathcal{O}(|\Delta \mathbf{r}|^2) ,
\end{align}
and the surface wave becomes
\begin{align}
\zeta_{N}(\mathbf{r}) &\approx \frac{A}{2 \pi} \int^{\pi}_{-\pi} d\phi e^{i k_F  |\mathbf{r} - \mathbf{r}_N|  \sin \phi } \notag \\
&\times \sum\limits_{n=1}^N   e^{i k_F (N-n)  \Delta \mathbf{r}  \cdot \frac{\mathbf{r} - \mathbf{r}_N}{|\mathbf{r} - \mathbf{r}_N|}   \sin \phi }  e^{-(N - n) / M} , \\
&= \frac{A}{2 \pi} \int^{\pi}_{-\pi} d\phi e^{i k_F  |\mathbf{r} - \mathbf{r}_N|  \sin \phi } \notag \\
&\times \left[\frac{1 - e^{\left(i k_F  \Delta \mathbf{r}  \cdot \frac{\mathbf{r} - \mathbf{r}_N}{|\mathbf{r} - \mathbf{r}_N|}  \sin \phi  - 1/M \right) N}}{1-e^{(i k_F  \Delta \mathbf{r}   \cdot \frac{\mathbf{r} - \mathbf{r}_N}{|\mathbf{r} - \mathbf{r}_N| } \sin \phi  - 1/M)}} \right] .
\end{align}
For large $N$, $\exp\left[\left(i k_F  \Delta \mathbf{r}  \cdot \frac{\mathbf{r} - \mathbf{r}_N}{|\mathbf{r} - \mathbf{r}_N|}  \sin \phi  - 1/M \right) N\right]$ oscillates rapidly over the characteristic time $T_F$ and averages to zero.  We also note that $k_F  \Delta \mathbf{r}   \cdot (\mathbf{r} - \mathbf{r}_N) / |\mathbf{r} - \mathbf{r}_N|$ is small and therefore the denominator simplifies to,
\begin{align}
 (1-e^{i k_F  \Delta \mathbf{r}   \cdot \frac{\mathbf{r} - \mathbf{r}_N}{|\mathbf{r} - \mathbf{r}_N| } \sin \phi  - 1/M}) \approx 1/M ,
 \end{align}
giving,
\begin{align}
\zeta_{N}(\mathbf{r}) &\approx \tilde{A}\int^{\pi}_{-\pi} e^{i k_F  |\mathbf{r} - \mathbf{r}_N|  \sin \phi }d\phi = \tilde{A} J_0 \left( k_F  |\mathbf{r} - \mathbf{r}_N| \right) , \label{eqn:DiscreteSurfaceWave}
\end{align}
where $\tilde{A} = A M$ is the memory dependent amplitude after many bounces.

Finally, to establish an effective analogy with the spatially continuous motion of a quantum particle, we transfer from the discrete bounces modeled in Eq.~(\ref{eqn:DiscreteSurfaceWave}) into the continuous domain.  We are interested in total path lengths of approximately $40-120\,\mathrm{mm}$ which is much larger than the distance the droplet travels over one Faraday period, $\Delta r = v_d T_F \approx 0.4\,\mathrm{mm}$. We can therefore take a coarse grained approach and let $\ddot{\mathbf{r}}_{n} \rightarrow \ddot{\mathbf{r}}$, $t_N \rightarrow t$, $\mathbf{r}_N \rightarrow \mathbf{r}_d(t) =  (t/ T_F) \Delta \mathbf{r}$ and return to the continuous representation of the surface height from Eq.~(\ref{eqn:DiscreteForce}), $\zeta_{N}(\mathbf{r}) \rightarrow \zeta(\mathbf{r}, t)$, giving
\begin{align}
\zeta(\mathbf{r}, t) &\approx \tilde{A} J_0\left(k_F  |\mathbf{r} - \mathbf{r}_d(t)| \right) . \label{eqn:travelingWave}
\end{align}
With this continuous transformation we can now have a suitable surface wave that we can use as input for the numerical simulation of the single-slit diffraction and the double-slit interference experiments.

\section{Interference and diffraction with droplets}  \label{sec:slits}

Following Feynman we ``choose to examine a phenomenon which is impossible, absolutely impossible, to explain in any classical way, and which has in it the heart of quantum mechanics. In reality, it contains the only mystery.''~\cite{feynman2011feynman}.  He is, of course, speaking about the self-interference of a particle through one or more slits.  This well known phenomenon has a direct analogue in the bouncing droplet system as observed by Couder and Fort~\cite{bib:CouderSingle}.  We submit, however, that even though interference in the bouncing droplet system is analogous to the interference seen in quantum mechanics, the mechanism of that interference is quite different and care should be taken when considering to what extent the analogy is valid.  We can demonstrate the similarities and differences to quantum mechanics by examining the bouncing droplet system in light of the single and double-slit experiments.

We have from Eq.~(\ref{eqn:travelingWave}) a simple continuous surface wave traveling at the speed of the droplet, $\mathbf{v}_d$, which we can send through one or more slits.  We first consider the resultant surface wave on the other side of the slit while the droplet is still approaching, but far from, the slit configuration.  In this regime a simple application of Huygen's principle gives the wave pattern after a slit.  It is approximated by filling the slits with emitters of circular waves with the same wave number $k_F$ as the continuous surface wave excited by the droplet, Eq.~(\ref{eqn:travelingWave}) and an angular frequency $\omega_d = k_F |\mathbf{v}_d|$.  The shape of the surface wave emanating from one slit is
 \begin{align}
\zeta_{\textrm{eff}}(\mathbf{r}, t) &=  \frac{\tilde{A}}{L} \sum\limits_{l=0}^L \mathrm{Re} \left[e^{-i \omega_d t} H_0^{(1)} (k_F |\mathbf{r} - \mathbf{r}_l |)\right] ,  \label{eqn:huygens}  
\end{align}
where $\mathbf{r}_l$ for $0 < l< L$ spans the space of the slit, and $H_0^{(1)}$ is the outgoing Hankel function, the result of which can be seen in Fig.~\ref{fig:SingleSlitDensitiesAndTrajectories}(b).  

At this point we adopt an approach closer to Bohmian mechanics, namely we neglect the back action of the droplet on the surface wave near and after the slit configuration.  This corresponds to being in a high memory regime in which the surface wave built from many previous impacts of the droplet overshadows contributions to the surface wave due to a single impact of the droplet.  In this high memory regime the wave after the slits will still take the relatively simple form of Eq.~(\ref{eqn:huygens}) until the accruing contributions from the droplet begin to have a significant effect in comparison.  While we work in the high memory regime it is important not to ignore the memory effect altogether. Indeed, after the droplet passes through the slit the post-slit surface wave will begin to degrade since it is no longer being reinforced by waves coming through the slit.  This effect is addressed in our double-slit interference simulations in Sec.~\ref{sec:double}.

\subsection{Diffraction in the single-slit experiment} \label{sec:single}

\begin{figure}[ht]
 \includegraphics[width=1\linewidth]{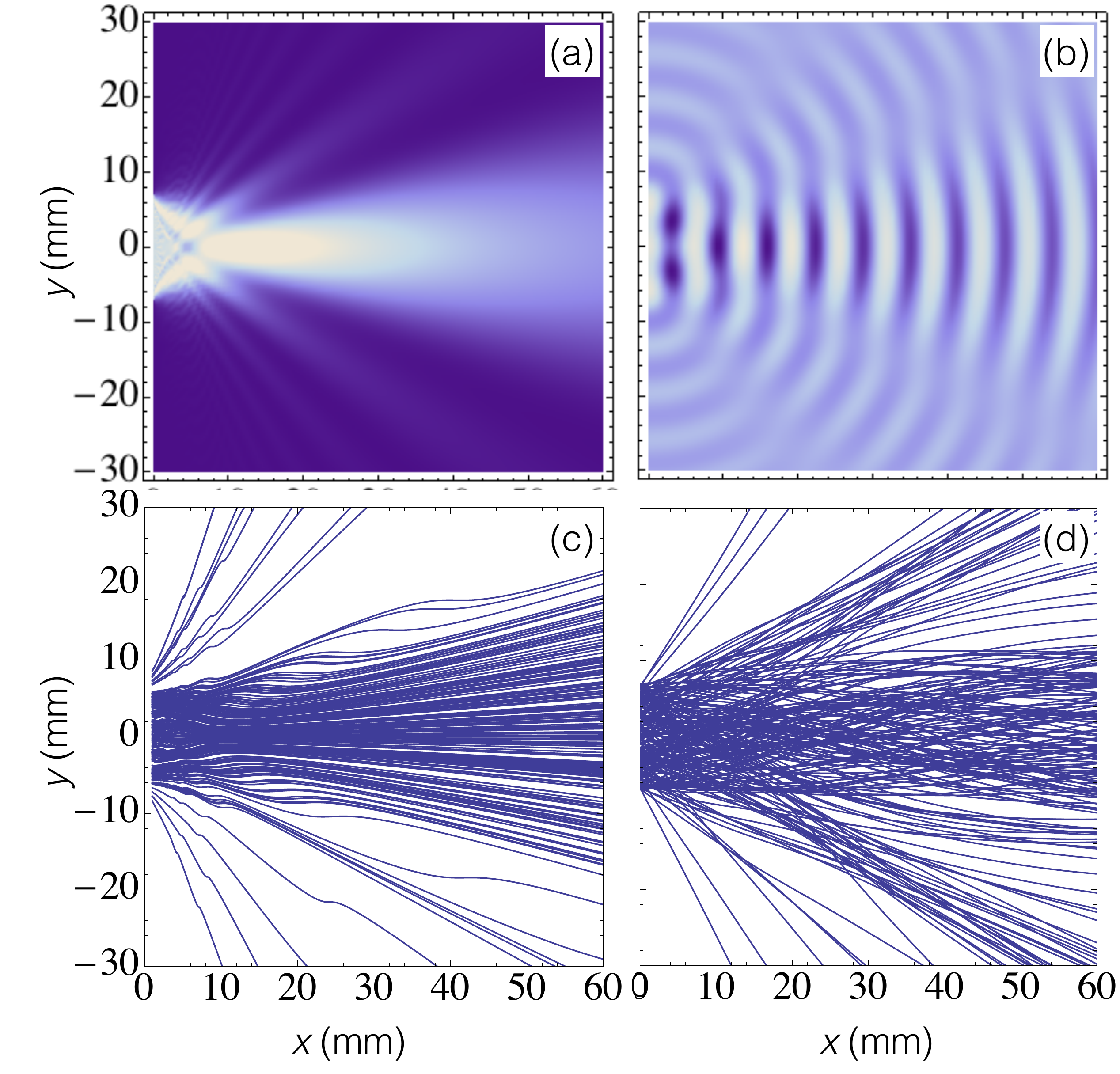}
\caption{(a) Density plot of the probability density of a quantum particle passing through a single slit.  (b) Density plot of the traveling surface wave of the bath after the droplet wave has passed through a slit.  (c) Trajectories due to Bohm's quantum potential, Eq.~(\ref{eqn:qmp}), of many quantum particles passing through a slit.  (d) Trajectories due to the hydrodynamic force, Eq.~(\ref{eqn:DiscreteForce}), of many droplets passing through a slit.  All plots have a slit width of $14\,\mathrm{mm}$, an initial velocity of magnitude $v_d$ with a cosine angular distribution, $k_F = 1\,\mathrm{rad/mm}$, $\omega_d = 10\,\mathrm{rad/s}$, and in the high memory regime with $M=1000$.}
\label{fig:SingleSlitDensitiesAndTrajectories}
\end{figure}

\begin{figure}[ht]
 \includegraphics[width=1\linewidth]{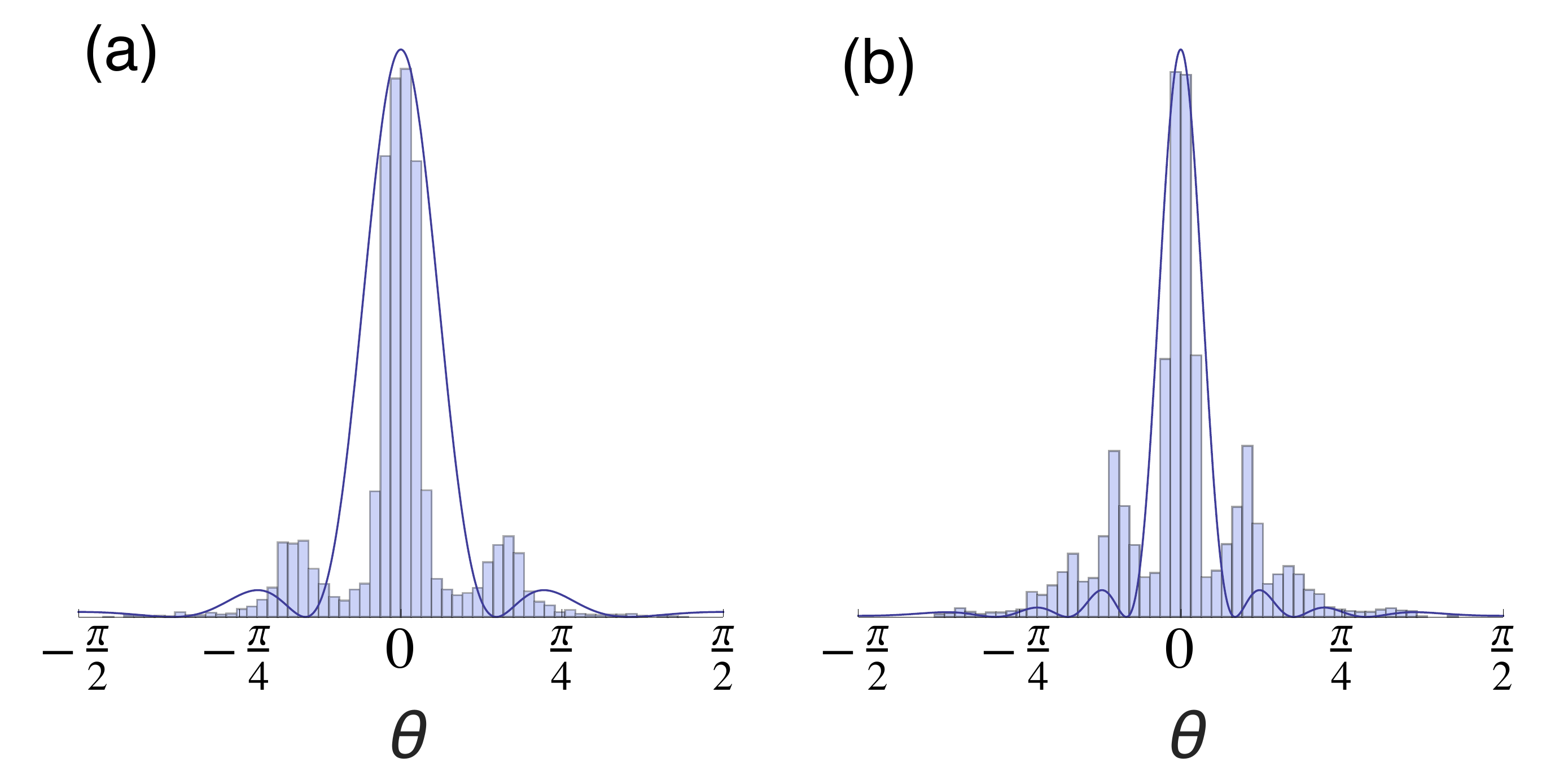}
\caption{Histograms showing the distribution of $4000$ angles, $\theta = \arctan(y/x)$, from the center of a single slit of width (a) $7\,\mathrm{mm}$ and (b) $12\,\mathrm{mm}$.  The angles are calculated from the positions of trajectories at a radius of $120\,\mathrm{mm}$ from the center of the slit.  The solid line is the diffraction pattern calculated from Bohmian trajectories eminating from a slit of identical width.  Note that minima predicted in the quantum case do not match those minima found from the droplet trajectories. Additionally, the probability of finding a droplet in the minima never reaches zero as it does for a particle in the quantum case.}
\label{fig:SingleSlitHistWidths}
\end{figure}

Using Eq.~(\ref{eqn:huygens}) with Eq.~(\ref{eqn:DiscreteForce}) we numerically compute the paths followed by many droplets with initial positions distributed uniformly along the slit, a $\cos \theta$ angular distribution for the angle $\theta$ between the direction of the initial velocity $\mathbf{v}_d$ of the droplet and the $x$-axis perpendicular to the center of the slit, with its magnitude $|\mathbf{v}_d|$ being fixed to the speed of a walking droplet in free space.  The results shown in Fig.~\ref{fig:SingleSlitDensitiesAndTrajectories}(d) can be contrasted to the trajectories of quantum particles shown in Fig.~\ref{fig:SingleSlitDensitiesAndTrajectories}(c), which are the result of Bohm's quantum force, Eq.~(\ref{eqn:quantforce}), and the the quantum probability density shown in Fig.~\ref{fig:SingleSlitDensitiesAndTrajectories}(a).  In the Bohmian case the initial velocities are derived from the probability current in the usual way~\cite{PhysRev.85.166,PhysRev.85.180}.  The resulting trajectories are fundamentally different but share some traits in the broad sense as can be seen from the histograms of angles shown in Fig.~\ref{fig:SingleSlitHistWidths}.  In particular it is easy to see that the quantum probability density is similar in form to the hydrodynamical results.  The minima predicted in the quantum case do not match those minima found from the droplet trajectories, which is not surprising as the quantum probability density is not analogous to the surface wave and care should therefore be taken with probabilistic analogies.  More importantly, however, the probability of finding a droplet in the minima never reaches zero as it does for a particle in the quantum case.  These discrepancies between the two models highlights a major difference between the hydrodynamic force and the quantum force.  The quantum force from Eq.~(\ref{eqn:quantforce}) is singular at points of zero probability, whereas the hydrodynamic force from Eq.~(\ref{eqn:DiscreteForce}) is not.  Consequently, there will always be a probability for a droplet to be found even in areas of very ``low'' probability.

\subsection{Interference in the double-slit experiment} \label{sec:double}

\begin{figure}[h!]
 \includegraphics[width=1\linewidth]{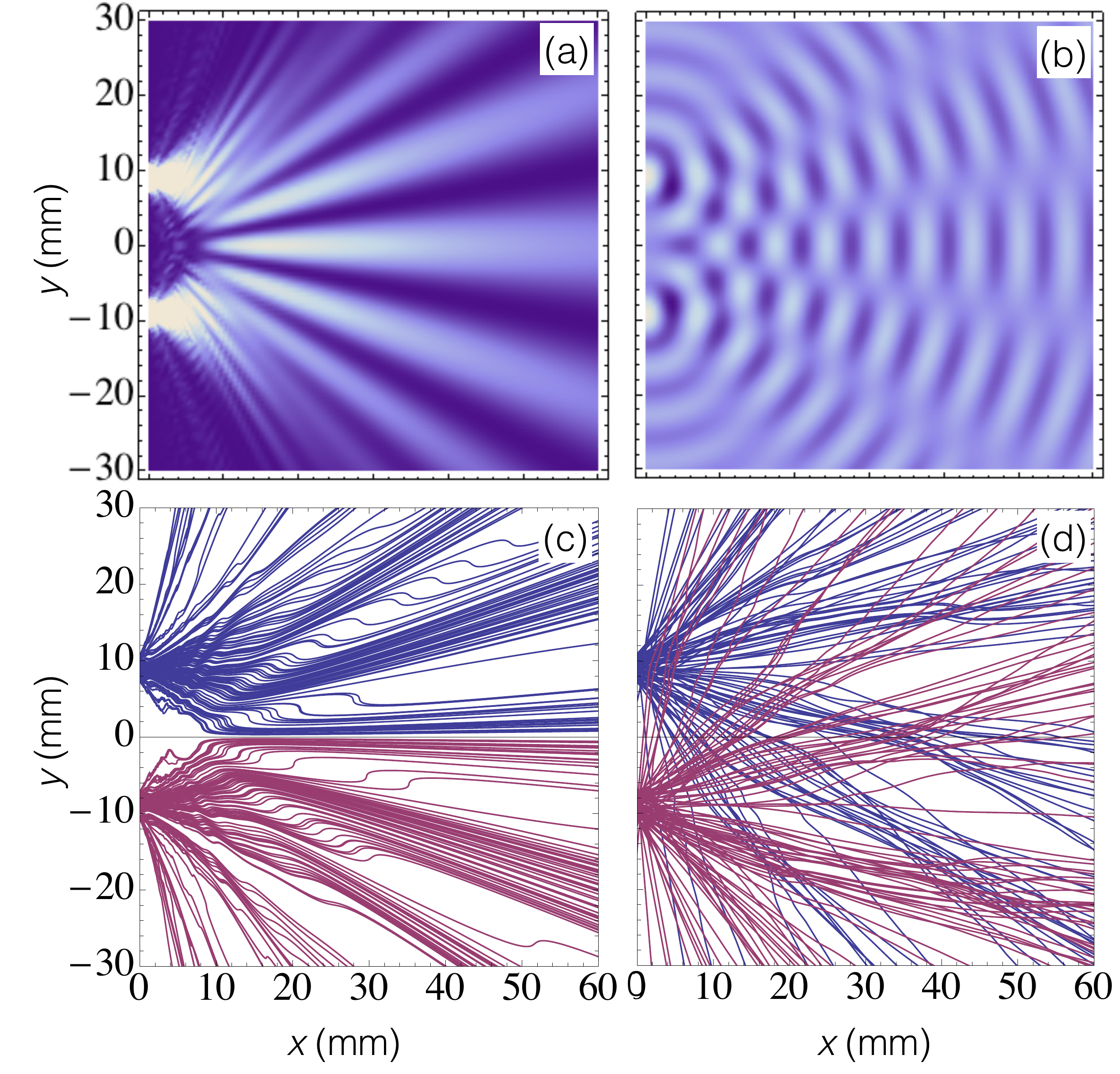}
\caption{(a) Density plot of the probability density of a quantum particle with an equal probability of passing through two slits.  (b) Density plot of the surface wave of the bath after the droplet wave has passed through the two slits.  (c) Trajectories due to Bohm's quantum potential, Eq.~(\ref{eqn:qmp}), of many quantum particles passing through the two slits.  (d) Trajectories due to the hydrodynamic force, Eq.~(\ref{eqn:DiscreteForce}), of many droplets passing through the two slits.  All plots have a slit width of $4\,\mathrm{mm}$ and a slit spacing of $18\,\mathrm{mm}$, equivalent parameters to the single-slit case and likewise being in the high memory regime with $M=1000$.}
\label{fig:DoubleSlitDensitiesAndTrajectories}
\end{figure}

\begin{figure}[h!]
 \includegraphics[width=1\linewidth]{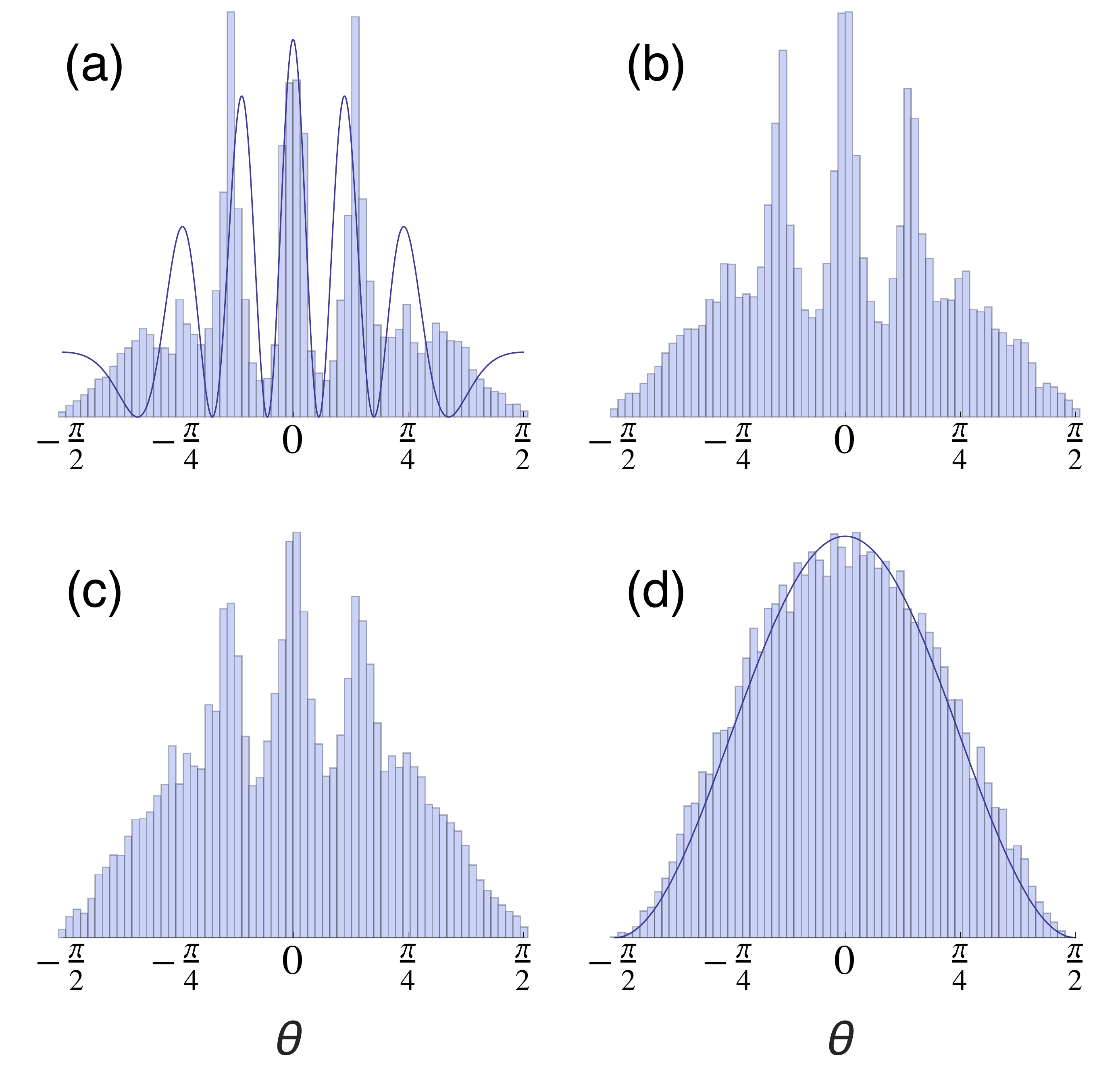}
\caption{Histograms showing the dependence on the memory parameter, $M$, of the distribution of $10000$ angles, $\theta$, from the center of two slits. The angle is calculated from the positions of trajectories at a radius of $40\,\mathrm{mm}$ from the midpoint between the slits.  (a) For $M=200$ the visibility, $V=0.80(4)$, calculated from Eq.~(\ref{eqn:visibility}) approaches the theoretical maximum value of $0.87$.  The solid line represents the theoretical interference pattern for a quantum double-slit experiment with identical slit widths and spacings. (b) $M=37.5$, $V=0.6(1)$ (c) $M=25$,$V=0.4(1)$ (d) At very low memory, $M=1$, the pattern reverts to the initial cosine angular distribution shown by the solid line and the visibility is zero.  The slit width, spacing, and parameters are the same as in Fig.~\ref{fig:DoubleSlitDensitiesAndTrajectories}. }
\label{fig:DoubleSlitVisibilityHistograms}
\end{figure}

\begin{figure}[h!]
\includegraphics[width=1\linewidth]{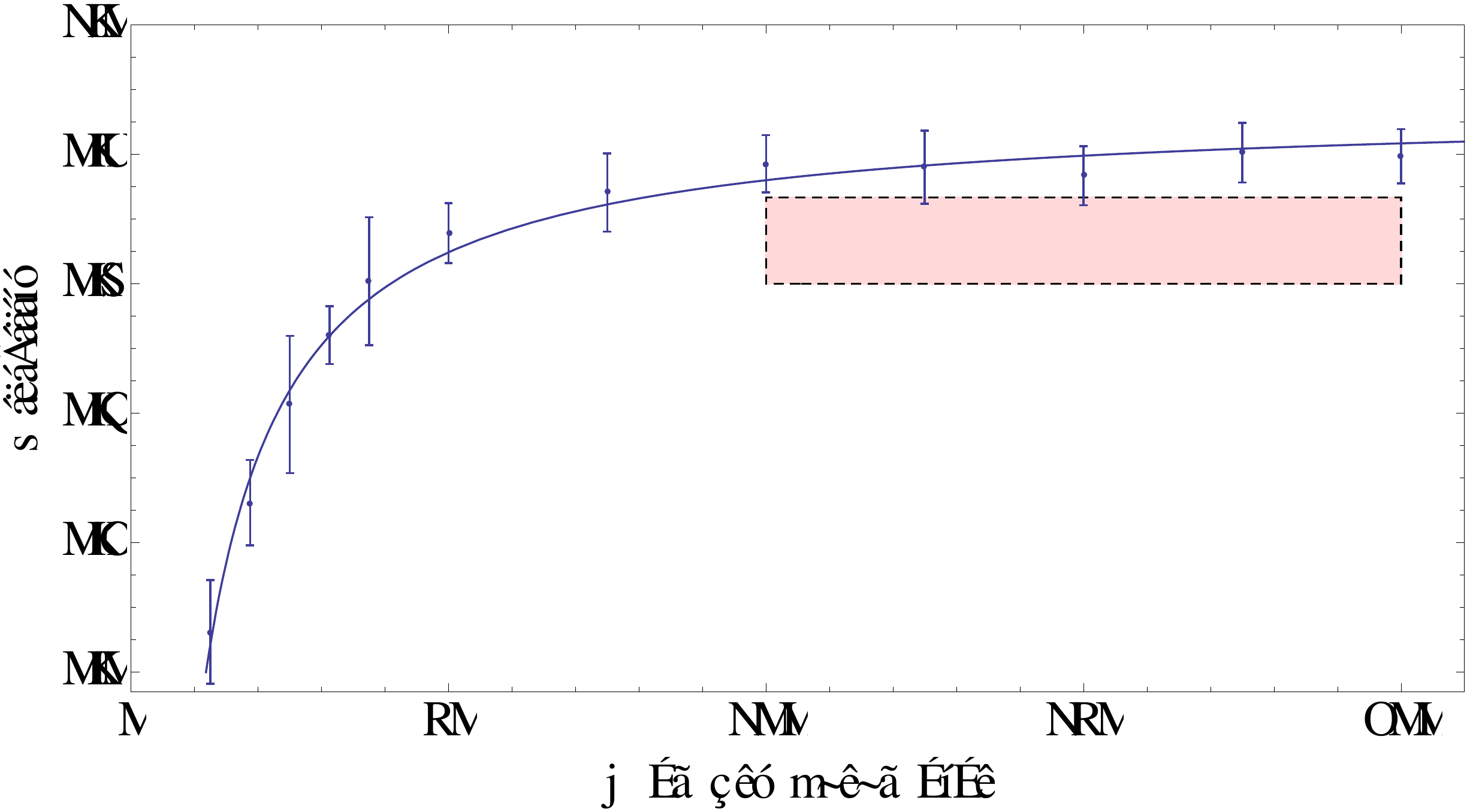}
\caption{The visibility, $V$, of the double-slit interference pattern as a function of the memory parameter, $M$.  The values for the visibilities and their error bars (plus or minus one standard deviation) are calculated from $10$ histograms with $1000$ trajectories each.  The solid curve is the heuristic function $V \approx V_{max}[1- f(M)]/[1+f(M)]$ with $f(M) = b/(M - c)$ best fit by the parameters $V_{max}=0.87(4)$, $b=7(2)$, $c = 5(2)$.  This fit gives a theoretical maximum visibility of $0.87$ and a minimum memory requirement on the order of $12$ to observe an interference pattern.  The shaded square (red) is the visibility estimated from Couder and Fort experiment~\cite{bib:CouderSingle} with parameters provided from private correspondence~\cite{CouderComm}.  The slit width, spacing, parameters, and sampling radius are the same as in Fig.~\ref{fig:DoubleSlitVisibilityHistograms}.}
\label{fig:VisibilityPlot}
\end{figure}

We now compute the trajectories for the double-slit experiment, Fig.~\ref{fig:DoubleSlitDensitiesAndTrajectories}(b), using the same parameters as for the single slit above.  The results are shown in Fig.~\ref{fig:DoubleSlitDensitiesAndTrajectories}(d) and can be contrasted against the Bohmian trajectories in Fig.~\ref{fig:DoubleSlitDensitiesAndTrajectories}(c) calculated from the quantum probability density shown in Fig.~\ref{fig:DoubleSlitDensitiesAndTrajectories}(a).  Here we note a striking contrast between the trajectories in the bouncing droplet system and those resulting from Bohmian mechanics.  One of the tenants of Bohmian trajectories is that the trajectories are forbidden to cross each other, and in the double-slit experiment the trajectories from each slit will not cross the center line, but it is obvious that the trajectories in Fig.~\ref{fig:DoubleSlitDensitiesAndTrajectories}(d) have no such reluctance to do so.  However, if we look at the histogram of angles given by Fig.~\ref{fig:DoubleSlitVisibilityHistograms}(a) we note that, as in the single-slit case, the interference pattern that builds gives similar results to that of the quantum case.

To further make comparisons with quantum mechanics we can look at the visibility $V$ of the central peak, defined by
\begin{align}
V = \frac{I_{\mathrm{max}}-I_{\mathrm{min}}}{I_{\mathrm{max}}+I_{\mathrm{min}}} , \label{eqn:visibility}
\end{align}
where $I_{max}$ is the height of the central peak and $I_{min}$ is the height of the first minima~\cite{PhysRevA.71.042103,ob:bornwolf,visibility3}.  In quantum mechanics the visibility diminishes as the``which path'' information increases, which has the effect of causing the quantum particle (as defined by the wave function) to behave more like a classical particle.  With increasing which path information the probability density becomes more dependent on a single slit.   Consequently the observed interference pattern becomes less pronounced as it is the wave function arising from \emph{both} slits that gives rise to the pattern.  We can make an analogy in the bouncing droplet system with the memory parameter being analogous to the which path information.  The lower the memory parameter is, the weaker the accumulated surface wave and the weaker the force on the droplet.  With a weaker surface wave the trajectory followed by a droplet tends to follow a straight line and therefore behaves more a classical free particle.  Plotting the histogram for several values of the memory parameter, Fig.~\ref{fig:DoubleSlitVisibilityHistograms}, we note that this is indeed the case and the visibility decreases as the memory decreases.

Plotting the observed visibilities for many values of the memory parameter as in Fig.~\ref{fig:VisibilityPlot} we see that interference fringes are visible for experimentally accessible values of the memory parameter, namely we have a visibility of $0.5$ at $M \approx 40$.  The visibility will increase with increasing memory but never reach one, in contrast to quantum particles with a which path information of zero.  This, as discussed in Sec.~\ref{sec:single}, is due to the difference between the quantum and the hydrodynamic force equation.  Unlike the quantum case there is never an area of the surface wave in which a particle will have zero probability of being encountered.  There is also a minimal amount of memory required for interference fringes to appear at all, in contrast to quantum interference in which the the visibility will smoothly approach zero as the which path information approaches one.  We note that while we still require the system to be in the high memory regime to build up the post-slit surface wave responsible for the single droplet interference, we have not imposed such a restriction after the droplet has passed the slit.

\section{Conclusion} \label{sec:conclusion}

In summary, we have numerically investigated single-slit diffraction and double-slit interference using droplets and compared the results to the predictions of the de Broglie-Bohm interpretation of quantum mechanics.  In order to be as close to this quantum interpretation as possible we have utilized a hydrodynamic model that neglects back-action of the droplet onto the surface wave after the droplet has passed the slit configuration, which corresponds to being in a high memory regime.  By way of Monte Carlo simulations based upon this model, we obtained diffraction and interference patterns in the angular probability distributions of the droplet trajectories that strongly resemble those reported by Couder and Fort~\cite{bib:CouderSingle} and that reflect a striking resemblance to quantum single-slit diffraction and double-slit interference on a phenomenological level.  However, we also identified evident differences from quantum mechanics, such as an imperfect fringe visibility which would not be present for a quantum particle.  These differences are ultimately traced back to the fundamentally different nature of the Bohmian force upon a quantum particle as compared to the force that the surface wave exerts upon a droplet.  In view of this, it is not obvious to what extent the present classical analogy of quantum wave-particle duality can be maintained in more complex situations involving, e.g., more than one droplet.  Further theoretical and experimental studies are clearly required to address this issue.
\begin{acknowledgments}
This work was financially supported by the Actions de Recherches Concert\'ees (ARC) of the Belgium Wallonia-Brussels Federation under contract No.~12-17/02.  We would also like to thank Yves Couder for discussions and supplementary experimental information.
\end{acknowledgments}

\end{document}